# Secret Sharing With Trusted Third Parties Using Piggy Bank Protocol

Adnan Memon

**Abstract**

This paper presents a new scheme to distribute secret shares using two trusted third parties to increase security and eliminate the dependency on single trusted third party. This protocol for communication between a device and two trusted third parties uses the piggy bank cryptographic paradigm. We also present a protocol to give law enforcing agencies access to sensitive information present on a cell phone or a device using secret sharing scheme. The ideas for classical systems may also be applied to quantum schemes.

**Introduction**

A trusted third party (TTP) can be defined as an entity which is trusted by both parties to facilitate communication. Imagine a scenario in which two parties completely trust the TTP and use it to exchange secrets, but it is compromised or it becomes dishonest, then the system's security will be completely broken. Thus, relying on a single trusted party is risky [1]. We can use two TTPs to reduce the risk and increase security, and in this paper we describe a protocol to distribute the secret shares using two TTPs to eliminate the dependency on single trusted third party in case it is compromised or becomes malicious.

Imagine a cell phone or a device with sensitive information which may be a threat to national security is in the custody of law enforcing agency but they cannot access it because it is password protected and the data wipes out automatically if the wrong password is entered few times. The cell phone or device owner company has also refused to provide any backdoor for information to be accessed stating the reason that the backdoor may come in wrong hands and hence the security of phones or devices in someone's physical ownership may be compromised. This in fact was what happened in the case between FBI and Apple a few months ago.

Keeping such scenarios in mind, the use of secret sharing scheme becomes logical. In this paper, we present a protocol in which the cell phone data or device data is accessible to law enforcing agencies using secret sharing scheme and two trusted third parties. It eliminates the dependency on cell phone or device owned company to access the information.

## Background: Data Partitioning Scheme

There have been several secret sharing schemes proposed in which the secret is partitioned and shared to various distributed servers. Shamir's secret sharing scheme [2] is based on polynomial interpolation and maps the data on y-axis whereas another data partitioning scheme [3] is based on the roots of a polynomial on the x-axis.

For background, consider the specific (k,n) threshold scheme based on polynomial roots [3] where k=2 and n=3. The data is partitioned into three pieces and stored on three servers (Fig. 1). Data reconstruction requires access to at least two out of three servers. This scheme may be called as (2,3) threshold scheme. The servers are chosen in such a way that only one piece of data goes to each server and this will secure the user's data from being accessed by any of them until at least two of them agree and combine their pieces to recreate the original data.

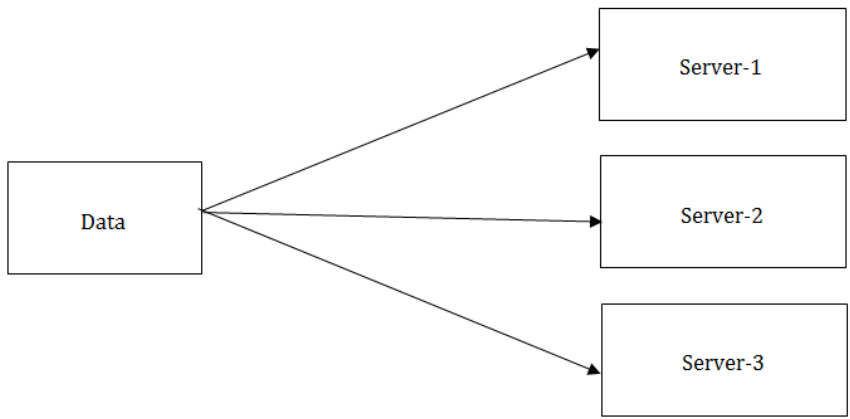

Fig. 1. Data Partitioning Process

The data is partitioned into three pieces using the fundamental theorem of algebra which states that every equation of degree three has three roots. Each of the partition is stored on a different server. To create n=3 partitions, first we create k=2 partitions. For this, we consider a 2nd order equation:

$$x^2 + a_1 x + a_0 = 0 \qquad (1)$$

Eq. (1) has 2 roots denoted by $\{r_1, r_2\} \subseteq \{set\ of\ complex\ numbers\}$. It can be rewritten as:

$$(x - r_1)(x - r_2) = 0 \qquad (2)$$

In cryptography, we may use finite field $\mathbb{Z}_p$ where p is a large prime. We can replace $a_0$ in (1) with data d $\in \mathbb{Z}_p$, then

$$x^2 + a_1 x + d = 0 \bmod p \quad \text{(d is input data)} \quad (3)$$

where $0 \leq a_i \leq p-1$ and $0 \leq d \leq p-1$ (One may also use – d in (3) in place of d)

This may be rewritten as:

$$(x - r_1)(x - r_2) = 0 \bmod p \quad (4)$$

where $1 \leq r_i \leq p-1$. The $r_i$ are the data partitions. The term d is independent of variable $x$ and therefore

$$r_1 \cdot r_2 = d \bmod p \quad (5)$$

There are two restrictions on choosing the coefficients:

1. Not all of them are simultaneously zero.
2. They should be so chosen such that a solution to the equation exists in $\mathbb{Z}_p$.

    **Step 1:** Choose randomly $r_1$ from a finite field $\mathbb{Z}_p$.

    **Step 2:** Compute $r_2 \equiv d \cdot (r_1)^{-1} \bmod p$.

    **Step 3:** Construct the 2nd degree polynomial: $p(2) = (x - r_1)(x - r_2) \bmod p = x^2 + a_1 x - d \bmod p$

    **Step 4:** The roots $r_1, r_2$ of the polynomial p(2) are the data partitions.

    **Step 5:** We generate 3 × 2 matrix A whose all rows are linearly independent. Vandermonde matrix of the following form is a simple way to choose such a matrix.

$$A = \begin{bmatrix} 1 & x_1 \\ 1 & x_2 \\ 1 & x_3 \end{bmatrix}$$

    **Step 6:** We create n=3 partitions by computing:

$$\begin{bmatrix} a_{11} & a_{12} \\ a_{21} & a_{22} \\ a_{31} & a_{32} \end{bmatrix} \begin{bmatrix} r_1 \\ r_2 \end{bmatrix} = \begin{bmatrix} c_1 \\ c_2 \\ c_3 \end{bmatrix}$$

    **Step 7:** The partitions are denoted by the pair $p_i = \{x_i, c_i\}$, $1 \leq i \leq n$.

**Example 1:** Let data d=12, prime p= 19, and k =2. We will use 2nd order equation to partition the data into two pieces, $x^2 + a_1 x - d = 0 \bmod p$. Assume $(x - r_1)(x - r_2) \bmod 19$. We choose one root randomly from the field, $r_1 = 13$. Therefore, $r_2 \equiv d \cdot (r_1)^{-1} \bmod 19 \equiv 12 \cdot (13)^{-1} \bmod 19 = 17$. The equation becomes $(x - 13)(x - 17) \equiv x^2 - 11x - 12 = 0 \bmod 19$. The coefficients $a_1 = 11$ and the partitions are 13 and 17.

## Data Reconstruction

**Step 1:** We construct $2\times 2$ matrix B by choosing the rows of matrix A corresponding to the given pairs $p_i$ and compute $B^{-1}$.

**Step 2:** We reconstruct the two (02) shares by evaluating:

$$\begin{bmatrix} r_1 \\ r_2 \end{bmatrix} = \begin{bmatrix} a_{m1} & a_{m2} \\ a_{i1} & a_{i2} \end{bmatrix}^{-1} \begin{bmatrix} c_1 \\ c_2 \end{bmatrix}$$

**Step 3:** Data d= $r_1 \cdot r_2 \bmod p$

## The Classical Piggy Bank Protocol for Communication between a TTP and the User

The piggy bank protocol [4] can be used between a TTP and the user as it can provide authentication, secures double-lock cryptography and counters MIM attack [5]. In the first step TTP sends an empty sealed piggy bank to the user. On receiving, user inserts the secret and the decryption key of a coded letter into the box together. The user also prepares a coded letter to be sent separately. TTP receives sealed piggy bank and the coded letter from user. The coded letter is sent to authenticate the contents of the sealed piggy bank box. It is also a secret and cannot be in plain text. TTP unlocks the secret within the letter with the decryption key present in the box. In this implementation of the piggy bank protocol for data, TTP obtains secret message (S) and key which is h(S).

> **Step 1.** TTP chooses a cryptographically strong random number R and the piggy bank transformation which is a one way transmission represented by f(R)= $R^e$ mod n, where n is a composite number with factors known only to TTP; e is the publicly known encryption exponent.
>
> **Step 2.** TTP sends f(R) to the user who multiplies it with its first secret S. The user sends $S(R^e) + K$ mod *n* to TTP in one communication and f(S)= $S^e$ mod *n* in another communication.
>
> **Step 3.** TTP uses its secret inverse transformation to first recover S and having found it can recover K.

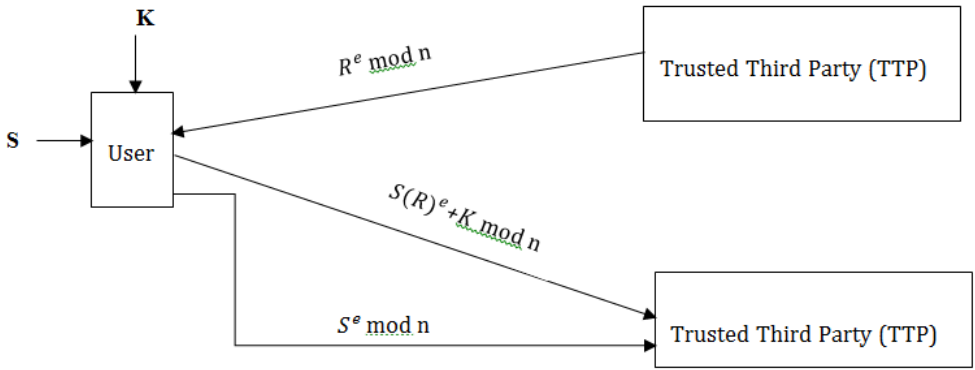
Fig. 2. The Piggy Bank Protocol for communication between a TTP and the user

**Example 1**:

Let composite number $n= 85$ and the public encrypting exponent is $e=5$. The secret decrypting algorithm be 13 since $5\times13 = 1 \bmod \varphi(85)$. Let TTP chooses random $R=19$ and computes $19^5 \bmod 85 = 49$ and send it to the user. Let user's random secrets be $S=6$ and $K=11$. User computes $6\times49+11=50$ and sends it in one communication and also $6^5 \bmod 85 = 41$ to TTP in another communication. TTP uses its secret decryption exponent 13 to recover S: $41^{13} \bmod 85 = 6$. Having found $S=6$, it computes $6\times49+K =50$, and recovers $K=11$.

## **Protocol for Distributing Secret Shares Involving Two Trusted Parties using Piggy Bank Protocol**

TTPs are widely employed for distributing secrets. This makes the use of TTPs more critical and make them a main target for distributed systems and cryptography research. Keeping in view such situation, we are describing a protocol to distribute the secret shares using two trusted third parties (TTPs) to increase security and to eliminate the dependency on single trusted third party in case it is compromised (Fig. 3). In this protocol, we are encrypting the secret share as well as the decryption key used to decrypt the secret share. Hence, we first decrypt the decryption key of secret share and then the secret share. The terms used in the protocol are explained as:

**Decryption key of the encrypted secret share decryption key:** This is used to decrypt the decryption key of (encrypted) secret share and it is not encrypted itself.

**Decryption key of the encrypted secret share:** This is used to decrypt the secret share.

### Step by Step Protocol

**Step 1:** Secret shares are created on user's device using some appropriate process.

**Step 2:** Only one unique secret share is encrypted and sent to TTP 1 along with the decryption key of the (encrypted) secret share decryption key. Each secret share is encrypted with a unique key.

**Step 3:** The decryption key of the (encrypted) secret share is encrypted and sent to TTP 2.

**Step 4:** TTP 1 sends unique recipient only one encrypted secret share and the decryption key of the (encrypted secret share decryption key.

**Step 5:** TTP 2 sends the recipient the encrypted secret share decryption key.

**Step 6:** Step 2 to Step 5 are repeated until every recipient receives its secret share.

**Step 7:** Every recipient can decrypt the secret share by first decrypting the encrypted secret share decryption key using the decryption key received from TTP 1 and then decrypting the encrypted secret share by the decryption key received from TTP 2.

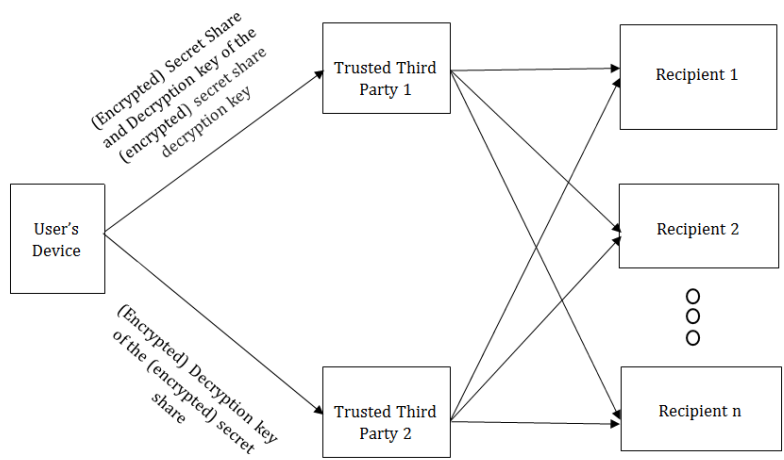

Fig. 3. Protocol for Distributing Secret Shares Involving Two Trusted Third Parties

### Application: Accessing Cell Phone/Device using Secret Sharing Scheme and Two TTPs

The above mentioned protocol can be applied in a variety of applications i.e. Cloud Computing. We are providing an application of the protocol keeping in view the scenarios presented in the Introduction. The data is partitioned into three pieces and stored on the servers of law enforcing agencies, judicial administration and cell phone/device owner company in such a way that only one piece of secret goes to each authority (Fig. 4). Data reconstruction requires access to at least 2 out of 3 servers.

Law enforcing agencies cannot access cell phone or device until one of other two combine their secret share with them. It is assumed that cell Phone or device owner companies would never share their secret share until it is the matter of national security because they would never want to lose the trust of people. In case, it is the matter of national security and cell phone or device owner company has denied to share its secret, the judicial administration can give its share to law enforcing agencies and access the cell phone or device. The number of servers can be *n* and any (k,n) secret share can be used to implement this. The decision to select authorities whom the secrets are sent is also on the will of the implementation authority.

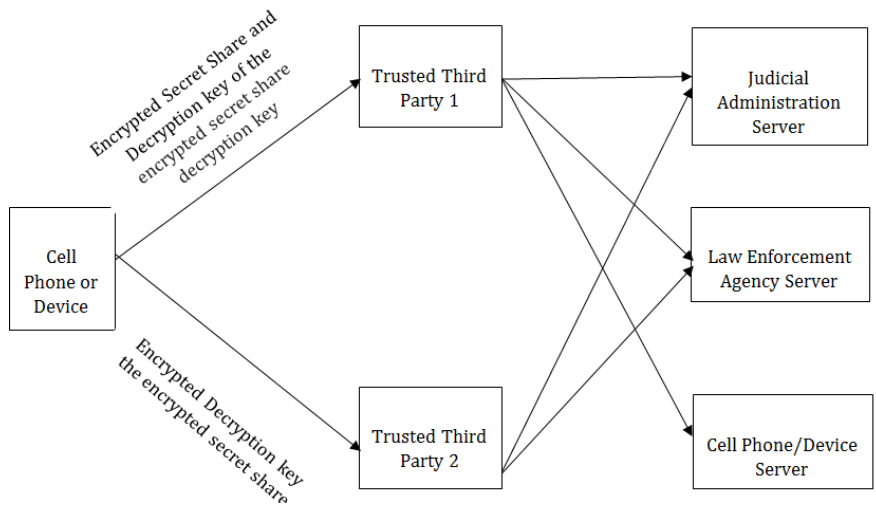

Fig. 4. Protocol for Accessing Cell Phone/Device using Secret Sharing Scheme and Two TTPs

## **Quantum Secret Sharing**

Now we come to quantum cryptography [6]. The concept of quantum secret sharing has been investigated and different schemes have been proposed. The procedure in which a message is split into various parts and all of them are required to read the message is described in [7]. A mechanism has also been shown to implement this procedure using GHZ states. The advantage of quantum case is the detection of eavesdropper because his presence will introduce errors. The concept of (k,n) threshold scheme is that a quantum state is divided into n shares and only k of those shares are needed to reconstruct the secret, but any set of k-1 or less shares reveal absolutely nothing about the secret [8]. Quantum no-cloning theorem puts a restriction on threshold schemes and requires n<2k and in our case n=3, k=2 and hence 3<2(2) = 3 <4, thus obeys the no-cloning theorem.

Sharing a three-state a qutrit (quantum trit) among three people [7]:

$$|0\rangle \rightarrow |000\rangle + |111\rangle + |222\rangle$$
$$|1\rangle \rightarrow |012\rangle + |120\rangle + |201\rangle$$
$$|2\rangle \rightarrow |021\rangle + |102\rangle + |210\rangle$$

Every person is holding either |0>, or |1>, or |2>, so has no information about the encoded state. But, any two people out of three can reconstruct the secret. For example, Alice and Bob, holding first two shares, then subtract the value of first from the second (mod 3). By doing this quantum mechanically, phase is also disentangled, thus reconstructing the state even if it is in a quantum superposition. This is an example of ((2, 3)) quantum threshold scheme in which any secret can be reconstructed by any two people but one person alone has no information.

**Quantum Piggy Bank Cryptography Protocol**

The Quantum mechanical version of piggy bank cryptography protocol is presented in [9]. The protocol can be as follows in the scenario of TTP and the user:

    **Step 1**. TTP applies the transformation $U_B$ on a random polarization state which is the cover state X, and sends n qubits of it to the user.

    **Step 2**. The user applies $U_A$ on the received *n* qubits to form $U_A U_B (X)$ and sends them back to TTP in one communication.

    **Step 3:** Somewhat later the user sends $U^+_A (Y)$ in another communication in *m* qubits where Y is the secret message bit (m << n).

    **Step 4**. TTP applies $U^+_B$ on the received bunch of *n* cover qubits and then performs tomography to get the transformation $U_A$ and then he applies this transformation on the second received communication to get secret message bit Y.

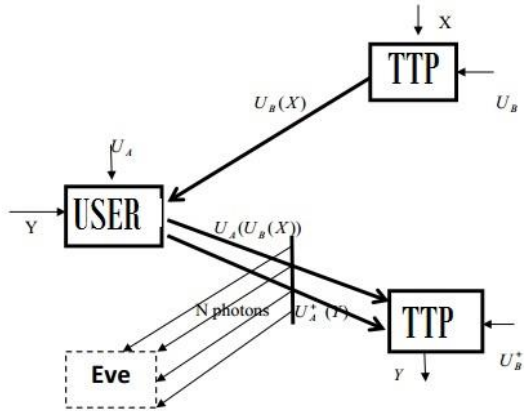

Fig.5 Quantum Piggy Bank Cryptography Protocol [9]

**Security Analysis**

In proposed protocol, we are using two TTPs and sending encrypted secret share via one TTP and its decryption key which is also encrypted via other TTP. This provides the confidentiality to the secret shares even in the case when one of the TTPs is compromised. If TTP 2 is compromised, the Eve will only have the encrypted secret share decryption key and not the secret shares. As the secret share decryption key is also encrypted, it provides another level of security to the secret shares and ultimately the secret. If TTP server 1 is compromised, the Eve will only have encrypted secret shares but not its decryption key and using better encryption technique and long decryption key will also make the brute force attack impossible for the Eve.

Use of Piggy Bank Protocol between an TTP and the user prevents Man in the Middle (MIM) attack if this protocol uses cryptographically strong RNG (Random Number Generation) algorithm or complex hash function. Both parties can introduce onetime pad or any other technique for authentication resource to recognize the misinformation from Eve, if she gets holds of the secret message. Piggy bank can also provide authentication.

Data partitioning scheme is implicitly secure as less than 2 shares cannot deduce any information about the secret. Eve has to access at least two servers to construct the secret.

**Conclusions**

This paper presented a new scheme to distribute secret shares using two trusted third parties to increase security and eliminate the dependency on single trusted third party. This protocol for communication between a device and two trusted third parties used the piggy bank cryptographic paradigm.